# Universal 2+1-Dimensional Plane Equations in General Relativity and Evolutions of Disk Nebula

*Yi-Fang Chang*
*Department of Physics, Yunnan University, Kunming, 650091, China*
(e-mail: yifangchang1030@hotmail.com)

ABSTRACT

The general relativity is the base for any exact evolutionary theory of large scale structures. We calculate the universal 2+1-dimensional plane equations of gravitational field in general relativity. Based on the equations, the evolutions of disk nebula are discussed. A system of nebula can form binary stars or single star for different conditions. While any simplified linear theory forms only a single star system. It is proved that the nonlinear interactions are very general, so the binary stars are also common.

At the present the general relativity with a precise formulation of the theory is the best astronomic theory for various large scale structures of space-time [1,2]. Since the general solutions of Einstein's field equations are very complex, so far, some exact solutions [1,2,3], including from the Schwarzschild solution, Eddington-Robertson solution, and various Kerr solutions to the gravitational lensing in metric theories of gravity [4], spherically symmetric space-times in massive gravity [5], and all static circularly symmetric perfect fluid solutions of 2+1 gravity [6]. They are mostly some static solutions.

On the other hand, the evolutions of various celestial bodies exist widely in the universe. In particular, astronomers observed that binary star systems are very common from 1989 [7,8]. The generality of binary star is proposed. Fact, Spyrou presented the results of a systematic study of relativistic celestial mechanics of binary stars in the post-Newtonian approximation (PNA) of general relativity [9], and proposed star models determined the inertial and rest masses of binary stars [10]. Itoh, et al., discussed the equation of motion for relativistic compact binaries with the strong field [11]. Alvi and Liu studied the dynamics of a cluster of collisionless particles orbiting a non-rotating black hole, which is part of a widely separated circular binary, and found that the most stable orbits are close to the companion's orbital plane and retrograde with respect to the companion's orbit [12]. Gu¨¨ro and Letelier studied binary systems around a black hole [13]. Hansen discussed the motion of the binary system composed of an oscillating and rotating coplanar dusty disk and a point-like object [14].

The most exact celestial evolutionary theory should be based on the general relativity. Wilson and Mathews [15,16] reported preliminary results obtained with a relativistic numerical evolution code. Their dynamical calculations suggest that the neutron stars may collapse to black holes prior to merger. Baumgarte, et al., studied the quasiequilibrium model on binary neutron stars in general relativity [17]. Then they presented a new numerical method for the construction of quasiequilibrium formulation of black hole-neutron star binaries, and solved the constraint equations of general relativity, and solved these coupled equations in the background metric of a Kerr-Schild black hole, which accounts for the neutron star's black hole companion [18]. Taniguchi, et al., presented new sequences of general relativistic, quasiequilibrium black hole-neutron star binaries. They solved the gravitational field equations coupled to the equations of relativistic hydrostatic equilibrium for a perfect fluid [19]. Shibata, et al., presented a new implementation for magnetohydrodynamics (MHD) simulations in full general relativity (involving dynamical space-times), and performed numerical simulations for standard test problems in relativistic MHD, including special relativistic magnetized shocks, general relativistic magnetized Bondi flow in stationary space-time, and a long-term evolution for self-gravitating system composed of a neutron star and a magnetized disk in full general relativity [20]. Faber, et al., calculated the dynamical evolutions of merging black hole-neutron star binaries that construct the combined black hole-neutron star space-time in a general relativistic framework. They treated the metric in the conformal flatness approximation, and assumed that the black hole mass is sufficiently large compared to that of the neutron star so that the black hole remains fixed in space [21].

One of more successful theories of the formation of binary stars is the fragmentation proposed by Boss, et al.[22-24], which supposes that binary stars are born during the protostellar collapse phase under their own gravity. Using computer simulation they obtained that an initial spherical cloud in rapidly rotation collapses and flattens to a disk, which later fragments into a binary system. Based on the basic equations of a rotating disk on the nebula, we use the qualitative analysis theory of nonlinear equation and obtain a nonlinear dynamical model of formation of binary stars [25]. Under certain conditions a pair of singular points results in the course of evolution, which corresponds to the binary stars. Under other conditions these equations give a single central point, which corresponds to a single star. This new method and model may be extended and developed. Steinitz and Farbiash established the correlation between the spins (rotational velocities) in binaries, and show that the degree of spin correlation is independent of the components' separation. Such a result might be related for example to the nonlinear model for the formation of binary stars from a nebula [26].

So far, the general relativity is mainly applied to neutron star-neutron star binaries, or black hole-neutron star binaries. Wilson, et al., chose the 3-metric to be conformally flat [15,16]. In Cartesian coordinates the line element can be written [17]

$$ds^2 = -\alpha^2 dt^2 + \Psi^4 \delta_{ij}(dx^i - \omega^i dt)(dx^j - \omega^j dt). \tag{1}$$

In more general cases, for a flatten disk nebula we take a universal 2+1-dimensional plane metric into diagonal form with 2+1-paramers

$$ds^2 = g_{ik}dx^i dx^k = e^\nu dx_0^2 - e^\lambda dr^2 - f(t,r,\theta)d\theta^2. \tag{2}$$

where $\nu = \nu(t,r,\theta)$ and $\lambda = \lambda(t,r,\theta)$. Denoting by $x^0, x^1, x^2$, respectively, the (2+1)-dimensional plane coordinates ct, r, $\theta$, we have for the nonzero components of the metric tensor the expressions [27]:

$$g_{00} = e^\nu, g_{11} = -e^\lambda, g_{22} = -f(t,r,\theta), \tag{3}$$

and $\quad g^{00} = e^{-\nu}, g^{11} = -e^{-\lambda}, g^{22} = -f^{-1}. \tag{4}$

With these values the calculation leads to the following Christoffel symbols:

$$\Gamma^0_{00} = \frac{1}{2}\dot{v}, \Gamma^1_{00} = \frac{1}{2}e^{v-\lambda}v', \Gamma^0_{01} = \frac{1}{2}v', \tag{5a}$$

$$\Gamma^2_{00} = \frac{1}{2f}e^v\frac{\partial v}{\partial \theta}, \Gamma^0_{02} = \frac{1}{2}\frac{\partial v}{\partial \theta}, \Gamma^1_{01} = \frac{1}{2}\dot{\lambda}, \tag{5b}$$

$$\Gamma^0_{11} = \frac{1}{2}e^{\lambda-v}\dot{\lambda}, \Gamma^2_{02} = \frac{1}{2f}\dot{f}, \Gamma^0_{22} = \frac{1}{2}e^{-v}\dot{f}, \tag{5c}$$

$$\Gamma^1_{11} = \frac{1}{2}\lambda', \Gamma^2_{22} = \frac{1}{2f}\frac{\partial f}{\partial \theta}, \Gamma^1_{12} = \frac{1}{2}\frac{\partial \lambda}{\partial \theta}, \tag{5d}$$

$$\Gamma^2_{11} = -\frac{1}{2f}e^\lambda\frac{\partial \lambda}{\partial \theta}, \Gamma^2_{12} = \frac{1}{2f}f', \Gamma^1_{22} = -\frac{1}{2}e^{-\lambda}f'. \tag{5e}$$

Here the prime means differentiation with respect to r, while a dot on a symbol means differentiation with respect to ct. All other components are zero. Therefore, we may derive the components of the field equations:

$$G^0_0 = -\frac{1}{4f}e^{-v}\dot{f}\dot{\lambda} + \frac{1}{4f}e^{-\lambda}[2f''-f'\lambda'-\frac{1}{f}(f')^2] +$$
$$\frac{1}{4f}[2\frac{\partial^2 \lambda}{\partial \theta^2}+(\frac{\partial \lambda}{\partial \theta})^2 - \frac{1}{f}\frac{\partial f}{\partial \theta}\frac{\partial \lambda}{\partial \theta}] = -\frac{8\pi k}{c^4}T^0_0, \tag{6a}$$

$$G^1_1 = \frac{1}{4f}e^{-v}[\dot{f}\dot{v}-2\ddot{f}+\frac{1}{f}\dot{f}^2] + \frac{1}{4f}e^{-\lambda}f'v' +$$
$$\frac{1}{4f}[2\frac{\partial^2 v}{\partial \theta^2}+(\frac{\partial v}{\partial \theta})^2 - \frac{1}{f}\frac{\partial f}{\partial \theta}\frac{\partial v}{\partial \theta}] = -\frac{8\pi k}{c^4}T^1_1, \tag{6b}$$

$$G^2_2 = \frac{1}{4}e^{-v}[\dot{v}\dot{\lambda}-\dot{\lambda}^2-2\ddot{\lambda}] + \frac{1}{4}e^{-\lambda}[2v''-v'\lambda'+(v')^2] +$$
$$\frac{1}{4f}\frac{\partial v}{\partial \theta}\frac{\partial \lambda}{\partial \theta} = -\frac{8\pi k}{c^4}T^2_2, \tag{6c}$$

$$G^1_0 = \frac{1}{4f}e^{-\lambda}[\dot{f}v'+f'\dot{\lambda}+\frac{1}{f}\dot{f}f'-2\dot{f}'] = -\frac{8\pi k}{c^4}T^1_0, \tag{7a}$$

$$G^2_0 = \frac{1}{4f}[-2\frac{\partial\dot{\lambda}}{\partial \theta}+\dot{\lambda}\frac{\partial(v-\lambda0)}{\partial \theta}+\frac{1}{f}\dot{f}\frac{\partial \lambda}{\partial \theta}] = -\frac{8\pi k}{c^4}T^2_0, \tag{7b}$$

$$G^1_2 = \frac{1}{4}e^{-\lambda}[-2\frac{\partial v'}{\partial \theta}-v'\frac{\partial(v-\lambda)}{\partial \theta}+\frac{1}{f}f'\frac{\partial v}{\partial \theta}] = -\frac{8\pi k}{c^4}T^1_2. \tag{7c}$$

These equations may possess various solutions.

The energy-momentum tensor of perfect fluid is [3]

$$T_{ab} = (\mu + p)u_a u_b + pg_{ab}. \tag{8}$$

Assume that $\dfrac{G_0^0}{G_1^1} = \dfrac{T_0^0}{T_1^1} = b$ and $\dfrac{G_0^0}{G_2^2} = \dfrac{T_0^0}{T_2^2} = d$, so

$$-\frac{1}{f}\dot{f}\dot{\lambda} = \frac{1}{f}b_1[\dot{f}\dot{v} - 2\ddot{f} + \frac{1}{f}\dot{f}^2] = d_1(\dot{v}\dot{\lambda} - \dot{\lambda}^2 - 2\ddot{\lambda}), \tag{9a}$$

$$\frac{1}{f}[2f'' - f'\lambda' - \frac{1}{f}(f')^2] = \frac{1}{f}b_2 f'v' = d_2(2v'' + v'^2 - v'\lambda'), \tag{9b}$$

$$[2\frac{\partial^2 \lambda}{\partial \theta^2} + (\frac{\partial \lambda}{\partial \theta})^2 - \frac{1}{f}\frac{\partial f}{\partial \theta}\frac{\partial \lambda}{\partial \theta}] = b_3[2\frac{\partial^2 v}{\partial \theta^2} + (\frac{\partial v}{\partial \theta})^2 -$$
$$\frac{1}{f}\frac{\partial f}{\partial \theta}\frac{\partial v}{\partial \theta}] = d_3 \frac{\partial v}{\partial \theta}\frac{\partial \lambda}{\partial \theta}. \tag{9c}$$

Here $b_1 + b_2 + b_3 = b$ and $d_1 + d_2 + d_3 = d$. For the evolutions of disk nebula, we consider a system only with time. Let $\dot{\lambda} = x, d(\ln f)/dt = \dot{f}/f = y$, the equations (9a) become

$$\frac{dx}{dt} = \frac{1}{2}[\frac{1}{d_1}xy + \dot{v}x - x^2], \tag{10a}$$

$$\frac{dy}{dt} = \frac{1}{2}[\frac{1}{b_1}xy + \dot{v}y - y^2], \tag{10b}$$

From dx/dt=0 and dy/dt=0 we derive four solutions:

$$(x, y) = (0,0), (0, \dot{v}), (\dot{v}, 0), (\frac{b_1(d_1 + 1)}{b_1 d_1 - 1}\dot{v}, \frac{d_1(b_1 + 1)}{b_1 d_1 - 1}\dot{v}). \tag{11}$$

Using a qualitative analysis theory of the nonlinear equations, the characteristic matrix of the equations (10) is

$$\begin{pmatrix} y/2d_1 + \dot{v}/2 - x & x/2d_1 \\ y/2b_1 & x/2b_1 + \dot{v}/2 - y \end{pmatrix}. \tag{12}$$

Its characteristic equation is

$$\lambda^2 - T\lambda + D = 0. \tag{13}$$

While $\Delta = T^2 - 4D$. For a singular point (0,0),

$$T_1 = \dot{v}, D_1 = \dot{v}^2/4 \geq 0, \Delta_1 = 0. \tag{14}$$

It is a critical nodal point. For a singular point $(0, \dot{v})$,

$$T_2 = \frac{1}{2d_1}\dot{v}, D_2 = -\frac{1}{4}(1 + \frac{1}{d_1})\dot{v}^2 \leq 0, \Delta_2 = (1 + \frac{1}{d_1} + \frac{1}{4d_1^2})\dot{v}^2 \geq 0. \tag{15}$$

It is a saddle point. For a singular point ($\dot{v}$,0),

$$T_3 = \frac{1}{2b_1}\dot{v}, D_3 = -\frac{1}{4}(1+\frac{1}{b_1})\dot{v}^2 \leq 0, \Delta_3 = (1+\frac{1}{b_1}+\frac{1}{4b_1^2})\dot{v}^2 \geq 0. \tag{16}$$

It is also a saddle point. For a singular point $(\frac{b_1(d_1+1)}{b_1 d_1 - 1}\dot{v}, \frac{d_1(b_1+1)}{b_1 d_1 - 1}\dot{v})$,

$$T_4 = -\frac{b_1 + d_1 + 2b_1 d_1}{2(b_1 d_1 - 1)}\dot{v}, D_4 = \frac{(b_1+1)(d_1+1)}{4(b_1 d_1 - 1)}\dot{v}^2,$$

$$\Delta_4 = \frac{1}{4(b_1 d_1 - 1)^2}[4(1+b_1+d_1)+(b_1+d_1)^2]\dot{v}^2. \tag{17}$$

In the usual case it is a nodal point. When $b_1 d_1 > 1$, i.e., $T<0$, the point is a stable sink, this system may form the binary stars. When $b_1 d_1 < 1$, i.e., $T>0$ and $D<0$, the point is a saddle point, this system forms a single star for (0,0) singularity. They seem to correspond to thin nebula and dense nebula, respectively.

When the singularity is a nodal point, $x = \dot{\lambda}, y = \dot{f}/f$ may be represented by $x_1 = ae^{ht}, y_1 = be^{kt}$ ($h<k<0$). So $\lambda = \int x dt = a_1 e^{ht} < 0, \ln f = \int y dt = b_1 e^{kt} < 0$ is also a nodal point.

The equations (6) are simplified to

$$\frac{1}{4f}e^{-v}\dot{f}\dot{\lambda} = \frac{8\pi k}{c^4}T_0^0 = aT_0^0, \tag{18a}$$

$$\frac{1}{4f}e^{-v}[\dot{f}\dot{v} - 2\ddot{f} + \frac{1}{f}\dot{f}^2] = -aT_1^1, \tag{18b}$$

$$\frac{1}{4}e^{-v}[\dot{v}\dot{\lambda} - \dot{\lambda}^2 - 2\ddot{\lambda}] = -aT_2^2. \tag{18c}$$

Then $xy = 4e^v aT_0^0$, and the equations (18b) and (18c) become

$$\frac{dx}{dt} = \frac{1}{2}\dot{v}x - \frac{1}{2}x^2 + 2e^v aT_2^2, \tag{19a}$$

$$\frac{dy}{dt} = \frac{1}{2}\dot{v}y - \frac{1}{2}y^2 + 2e^v aT_1^1, \tag{19b}$$

From dx/dt=0 and dy/dt=0 we derive solutions:

$$x_{1,2} = \frac{1}{2}(\dot{v} \pm \sqrt{\dot{v}^2 + 16e^v aT_2^2}) = \frac{1}{2}(\dot{v} \pm \sqrt{X}), \tag{20a}$$

$$y_{1,2} = \frac{1}{2}(\dot{v} \pm \sqrt{\dot{v}^2 + 16e^v aT_1^1}) = \frac{1}{2}(\dot{v} \pm \sqrt{Y}). \tag{20b}$$

Using a qualitative analysis theory, the characteristic matrix of the equations (19) is

$$\begin{pmatrix} \dot{v}/2 - x & 0 \\ 0 & \dot{v}/2 - y \end{pmatrix}. \tag{21}$$

Its characteristic equation is (13). In this case, for a singular point $(x_1, y_1)$,

$$T_1 = -\frac{1}{2}(\sqrt{X} + \sqrt{Y}) < 0, D_1 = \frac{1}{4}\sqrt{XY} > 0, \Delta_1 = \frac{1}{4}(\sqrt{X} - \sqrt{Y})^2 \geq 0. \tag{22a}$$

It is a nodal point, and is a stable sink. For a singular point $(x_1, y_2)$,

$$T_2 = -\frac{1}{2}(\sqrt{X} - \sqrt{Y}), D_2 = -\frac{1}{4}\sqrt{XY} < 0, \Delta_2 = \frac{1}{4}(\sqrt{X} + \sqrt{Y})^2 > 0. \tag{22b}$$

It is a saddle point. For a singular point $(x_2, y_1)$,

$$T_3 = \frac{1}{2}(\sqrt{X} - \sqrt{Y}), D_3 = -\frac{1}{4}\sqrt{XY} < 0, \Delta_3 = \frac{1}{4}(\sqrt{X} + \sqrt{Y})^2 > 0. \tag{22c}$$

It is also a saddle point. For a singular point $(x_2, y_2)$,

$$T_4 = \frac{1}{2}(\sqrt{X} + \sqrt{Y}) > 0, D_4 = \frac{1}{4}\sqrt{XY} > 0, \Delta_4 = \frac{1}{4}(\sqrt{X} - \sqrt{Y})^2 \geq 0. \tag{22d}$$

It is a nodal point, and is an unstable source. For the linear equations of simplified (10) or (19), only singularity (0,0) is a nodal point, and forms only a single star system.

Based on the universal 2+1-dimensional plane equations of gravitational field in general relativity, the evolutions of disk nebula are discussed. A system of nebula can form binary stars or single star for different conditions. Here the nonlinear interactions play an important role. They are necessary conditions of the formation of binary stars, but are not sufficient conditions.

Generally, any stable star should be a stable fixed point in the evolutionary process in astrophysics. Assume that an evolutionary equation is $y=f(x)$, which corresponds to an equation of the invariant point $x^*=f(x^*)$. Therefore, if $f(x)$ is a nonlinear n-order function, there will be possibly multiple (stable or unstable) invariant points. While $f(x)=ax+b$ is a linear function, there will be only an invariant point $x^*=b/(1-a)$, namely, it can form only a single star. Since the nonlinear interactions are very general, the binary and multiple stars are also common.